# THz detection with epitaxial graphene field effect transistors on silicon carbide

F. Bianco,[1, a)] D. Perenzoni,[2)] D. Convertino,[3)] S. L. De Bonis,[1, b)] D. Spirito,[1, c)] M. S. Vitiello,[1)] C. Coletti,[3)] M. Perenzoni,[2)] and A. Tredicucci[4)]

[1] *NEST, Istituto Nanoscienze-CNR and Scuola Normale Superiore, P.za S. Silvestro 12, 56127 Pisa (Italy)*

[2] *Fondazione Bruno Kessler (FBK), Povo, Italy*

[3] *CNI@NEST, Istituto Italiano di Tecnologia, P.za S. Silvestro 12, 56127 Pisa (Italy)*

[4] *NEST, Istituto Nanoscienze-CNR and Dipartimento di Fisica "E. Fermi", Università di Pisa, L.go Pontecorvo 3,*

We report on room temperature THz detection by means of antenna-coupled field effect transistors fabricated by using epitaxial graphene grown on silicon carbide substrate. The photoresponsivity is estimated in ~0.25 V/W and NEP ~ 80 nW/$\sqrt{\text{Hz}}$. Two independent detection mechanisms are found: plasma wave assisted-detection and thermoelectric effect, which is ascribed to the presence of junctions along the FET channel. The superposition of the calculated functional dependence of both the plasmonic and thermoelectric photovoltages on the gate bias qualitatively well reproduces the measured photovoltages. Additionally, the sign reversal of the measured photovoltage demonstrates the stronger contribution of the plasmonic detection compared to the thermoelectric mechanism. Although responsivity improvement is necessary, these results demonstrate that plasmonic detectors fabricated by epitaxial graphene on silicon carbide are potential candidates for fast large area imaging of macroscopic samples.

In addition to a variety of active and passive photonics devices[1] for the generation[2–4] and modulation[5–8] of terahertz (THz) radiation, graphene has been also successfully demonstrated as an excellent platform for low-cost room-temperature operation THz photodetection systems[1,9]. Generally, the photodetection is based on the conversion of incoming photons into an electrical signal. In the THz window, this conversion has been well achieved at room temperature by employing three main mechanisms have been reported: bolometric, thermoelectric (TE) and plasmonic effects. In bolometric detectors, the absorption of the incident photons in the substrate and graphene itself increases the detector temperature. This is associated to a temperature-induced shift of the graphene charge neutrality point (CNP) and thus to a conductivity variation of the graphene channel[10,11]. On the contrary, in thermoelectric and plasmonic detectors, a photovoltage is generated due to the presence of temperature gradient within the graphene channel (thermoelectric effect)[12] or to the rectification of the THz-induced ac current by means of the nonlinear response of the plasma waves excited into the channel of a graphene field effect transistor (GFET) (plasmonic effect)[13–17].

---

[a] Author to whom correspondence should be addressed. Electronic mail: federicabianco82@gmail.com
[b] Present address:
[c] Present address:

The large interest in reliable fast room-temperature photodetection is motivated by growing applications of the terahertz light in areas like spectroscopy, biomedical and security imaging. For imaging applications, particular attention is currently paid to develop sensitive, compact, portable, very fast imagers. Presently, room temperature detector arrays have been demonstrated with silicon CMOS[18] (hundreds μs-scale response time), pyroelectric[19] or microbolometeric[20] (response time in the millisecond time range) detector arrays. These detectors exhibit ~kV/W photoresponsity and low noise-equivalent-power (NEP < nW/$\sqrt{Hz}$). In this context, thanks to the atomic-scale thickness of the channel, the large scalability of the channel length up to very short sizes and the extremely high carrier mobility in graphene sheets, GFETs offer promise of high quality electronics performances[21]. Hence, in combination with plasmonic detection, GFETs are extremely interesting for realizing focal plane sensing devices with high responsivity and ultrafast response time. In order to fabricate a mosaic of closely spaced GFET-detector elements, large area graphene is strictly needed. THz detection by chemical vapor deposition (CVD) graphene-FET has been recently reported[16]. However, graphene films CVD-grown on an insulating substrate (like the one mostly used for transistors fabrication) are currently limited to discontinuous nanometer-sized islands[22]. Alternatively, CVD growth on copper foils is emerging as a powerful technique to obtain millimeter-sized single crystal graphene[23]. However, high-fidelity transfer process is required to maintain good crystalline integrity, long-distance continuity and cracks absence. Therefore, direct growth of graphene on the device substrate would be highly desirable. For this reason, epitaxial graphene represents a promising alternative. In fact, thanks to the graphitization of silicon carbide (SiC) by silicon (Si) sublimation, high quality graphene film can be directly grown on insulating SiC substrate in the form of large-diameter wafers.

In this study, we report on THz detection measurements at room temperature of an antenna coupled-GFET when using epitaxial graphene grown on the Si-face of SiC substrate. In this kind of detectors, the photo-induced voltage was generated by both plasma wave rectification and photothermoelectric effect due to the non-uniform charge profile across the FET channel.

The detector consisted of a log-periodic circular-toothed antenna (see inset of Fig. 1), whose lobes were connected to the source (S) and gate (G) of the GFET, while the drain (D) was a metal line. The 2.5 μm long and 2 μm wide channel was defined by etching the large area graphene with a polymethyl methacrylate (PMMA) mask via oxygen/argon reactive-ion etching. Antenna and metal contacts were patterned by electron beam lithography and thermally evaporated by a stack of chromium and gold with thickness of 5 nm and 100 nm, respectively. Finally, 35 nm-thick hafnium oxide was deposited by atomic layer deposition as gate dielectric for a top-gate configuration. The length of the gate ($L_g$) was 300 nm, corresponding to a gate-to-channel capacitance per unit area of $3.5\times10^{-7}$ F/cm$^2$. The FET channel was obtained by bilayer graphene grown on semi-insulating nominally on-axis-oriented 4H-SiC(0001) substrate using a resistively heated cold wall reactor (BM, Aixtron)[24]. Growth was performed at a temperature of 1390 °C and pressure of 780 mbar for 20 minutes. Assessment of the distribution of the number of layers on the sample was conducted via spatially-resolved Raman spectroscopy using a Renishaw InVia system equipped with a 532 nm laser[23]. The full width at half maximum (FWHM) of the 2D peak was mapped and correlated to AFM micrographs (Fig.1(a)). Trilayer inclusions were observed at the terrace edge as in Ref. 25 (Fig. 1(b)).

The detector was electrically characterized by two terminals-method, measuring the current ($I_{sd}$) as a function of gate voltage ($V_g$) while keeping source/drain voltage ($V_{sd}$) at 1 mV. The source/drain resistance ($R_{sd}$) varied from 14 to 17.5 kΩ with a maximum at gate voltage $V_{CNP}$ = 0.1 V. The observed p-doping can be ascribed to extrinsic contaminations (e.g. exposure to air and residual fabrication impurities), contrarily to the intrinsic electron-doping reported in as-growth epitaxial graphene on SiC due to the charged dangling bonds of the buffer layer[26]. Moreover, no evident signature of a band gap opening typical of bilayer graphene[27] was observed in the $R_{sd}$ curve. Tentatively, this can be caused by the fact that a large fluctuating disorder potential[28,29] and/or superposition of the band structures of bilayer with residual amounts of trilayer graphene[25] may mask the presence of the band gap. Mobility and residual carrier density were extracted by nonlinearly fitting the transfer characteristic[15], resulting in μ = 120 cm$^2$/Vs and $n_o$ = $1.7\times10^{12}$ cm$^{-2}$ in agreement with the values reported for epitaxial bilayer graphene[30]. Generally, the mobility in epitaxial graphene grown on Si terminated-SiC substrate is limited by the presence of the interfacial layer[26], that causes Coulomb scattering and optical and low-energy phonon scattering[31,32], and by the short-range scattering due to structural defects, step edges and thickness inhomogeneities[33]. In bilayer graphene the temperature dependence of the mobility has been found to be weak; as a consequence the electronic transport can be considered dominated by Coulomb and short range scattering mechanisms[30,34].

When the detection mechanism is based on the rectification of THz-induced ac current between source and drain, the carrier density and drift velocity are simultaneously modulated at the radiation frequency ω. As a result, the detector photoresponse has the form of a dc signal proportional to the incident power. Plasma waves are excited in the FET channel[13,17]; however, in the low frequency and long gate regime (ωτ ≪ 1, where τ is the carrier momentum relaxation time and gate length much longer than plasma decay length), only overdamped waves are excited, meaning that they decay before reaching the drain side of the channel. According to the hydrodynamic approach proposed by Dyaknov and Shur (DS)[13], the induced ac current can exist only up to a distance $l_{pl}$ from the source. This distance is quantified as $l_{pl} = s\sqrt{2\tau/\omega}$, where s is the plasma waves velocity (s ~ $10^6$ m/s [17]) and $\tau \sim \mu m^*/e$ the scattering time, with $m^* = 0.028 m_e$ the effective mass of bilayer graphene[35], $m_e$ the free electron mass, e the electron charge and µ the carrier mobility. Considering the mobility extracted from the FET resistance, the resulting scattering time was $\tau \sim 2$ fs, so that $\omega\tau \sim 0.003$, while the decay length $l_{pl} \sim 115$ nm < $L_g$. This means that a plasma wave-assisted detection in broadband overdamped regime was expected in our detector.

The photoresponsivity was characterized by illuminating the detector with a broadband THz source (WR2.8AMC, Virginia Diodes Inc.) that covered a spectral range from 230 to 375 GHz. The output power was frequency dependent and ranged from 0.5 to 1 mW (calibrated after the focusing lens). The source was modulated at 858 Hz with a fixed horizontal polarization. A lens was employed to focus the radiation and maximize the illumination of the detecting region. The photovoltage signal was then recorded by means of a lock-in amplifier (Signal Recovery mo. 7265 DSP) with an input impedance of 10 MΩ and variable internal gain.

Figure 1(c) shows the photovoltage (ΔU) measured as a function of the incident radiation frequency for beam polarization parallel (0 degree) and orthogonal (90 degree) to the antenna axis, while keeping $V_g = 1.2$ V and $V_{sd} = 0$ mV. As expected by dipole antenna operation, a polarization-sensitive signal was observed. The maximum photovoltage was recorded when the polarization was parallel to the antenna axis, thus ΔU versus light frequency exhibited a series of pronounced peaks, in accordance to the antenna geometry. Instead, a net photovoltage suppression occurred when the radiation was cross-polarized. In fact, only a clearer but weak feature was recognizable at 263 GHz.

Fixing the THz wave polarization at 0 degree, the photo-induced voltage was measured at the peak frequencies of Fig. 1(c) (i.e. 263 GHz, 295 GHz, 325 GHz and 353 GHz) by sweeping the gate voltage from negative to positive values (Fig. 2). All curves (ΔU ≡ $\Delta U_{mis}$) show nearly constant negative values at negative gate voltage, while an inverted V-shape with small sign switch at positive gate voltage. In the overdamped plasma wave regime, the solution of both Eulero equation (for the drift velocity) and continuity equation (for the carrier density)[13] predicts a detector photovoltage with gate voltage dependency as $\Delta U_{pl} = C\, \sigma^{-1} d\sigma/dV_g$. Here, σ is the source/drain conductance and C is the constant that takes into account the efficiency of radiation-to-antenna coupling and the FET impedance. The inset in Fig. 2 shows the $\Delta U_{pl}$ calculated as function of the gate voltage by using the measured FET conductance and assuming $C \sim 10^{-4}$ $V^2$, as estimated in similar GFET detectors[15]. By comparing the trend of $\Delta U_{mis}$ with $\Delta U_{pl}$, a general similarity between the curves was found, suggesting a detection mechanism based on DS model. Nevertheless, the model did not comprehensively describe the detector photoresponse. In fact, contrarily to what happens in $\Delta U_{pl}$, a clear change of the photovoltages sign when crossing the CNP voltage was not observed. Additionally, the photovoltages at CNP did not vanish like in $\Delta U_{pl}$, but showed a negative offset of few µV (Fig. 2) depending on the incoming frequency, as pointed out in Fig. 3(a). Interestingly, a three times smaller offset was found by rotating the light polarization at 90 degree at 263 GHz. The origin of this offset can be ascribed to the thermoelectric (TE) effect. In fact, a thermoelectric voltage ($\Delta U_{TE}$) can be generated by the presence of carrier density junctions created at the interface of ungated and gated regions[14,16,36–38]. In our detector, p-n/p-p junctions were formed across the FET channel depending on the applied gate bias. Owing to the antenna, the THz radiation is asymmetrically funneled onto the FET channel, inducing a local heating at the junction edge on the source side. In epitaxial graphene additional TE signal may arise also at the edges of graphene with different thickness, but this contribution is stronger at low temperature (T < $T_{room}$)[39]. The resulting nonequilibrium hot carriers distribution generates a temperature gradient ($\Delta T$) within the channel and thus a thermoelectric voltage according to $\Delta U_{TE} = (S_1 - S_2)\Delta T$, where $S_{1,2}$ are the thermopowers of the two regions with different carrier density. For estimating $\Delta U_{TE}$ in our detector, we used the Mott relation. This is a mesoscopic formula relating the thermopower S with the conductance σ achieved from the transport characteristic[40,41]

$$S = -\frac{\pi^2 k_B^2 T}{3e} \frac{1}{\sigma} \frac{d\sigma}{dE_F} , \qquad (1)$$

with $k_B$ the Boltzmann constant, T the sample temperature, e the electron charge and $E_F = k_B T_F$ the graphene Fermi energy and temperature ($T_F$). Eq. (1) is valid as long as $k_B T \ll E_F$ or when $T \leq 0.2 T_F$ independently on the dominant scattering mechanim[41]. Generally, it well reproduces the room temperature graphene thermopower measured at $V_g > V_{CNP}$[29], whereas it is valid near the CNP only if the carrier density is such that $k_B T \ll E_F$, like in low mobility graphene as a result of large residual charge fluctuations[42]. In our experiment $T/T_F \sim 0.3$ at CNP, thus a deviation from the Mott formula less than -30% is predicted both for unscreened charged impurities and short range scattering[41]. A larger deviation (~ -50%) is expected in the screening conditions[41,43]. Hence, without entering into details on the dominant scattering mechanism, Eq. (1) was used to roughly estimate the thermopower for gated ($S_g = S(V_g)$) and ungated ($S_{ug} = S(V_g=0)$) areas and the corresponding $\Delta S = S_g - S_{ug}$. At CNP $\Delta S \sim -2$ µV/K. Consequently, a $\Delta T$ between the junctions that ranges between about 0.7 and 1.5 K depending on the radiation frequency (Fig. 2) could be accountable of the non-zero measured photovoltage at CNP. Likely, the frequency dependence of $\Delta T$ is determined by the sensitivity of the antenna-radiation coupling on the frequency radiation. Instead, $\Delta T <$ 0.5 K was evaluated when rotating the incident radiation at 90 degree, pointing out the enhancement of the local heating by means of the antenna. Within this scenario, the thermoelectric voltage $\Delta U_{TE}$ could be estimated for all used radiation frequencies. As shown in Fig. 4(a) and 4(b), the calculated thermoelectric response contributes mainly at positive gate voltages and has a curve similar to the plasmonic photovoltage but with opposite sign. Thus, by considering the additional contribution of $\Delta U_{TE}$ to the plasmonic photoresponse, the overall behavior of the measured signal $\Delta U_{mis}$ could be qualitatively well reproduced. In fact, both the flat and V-shaped behavior at $V_g < 0$ and $V_g > 0$, respectively, may be explained by the superimposed functional dependence of $\Delta U_{TE}$ with $\Delta U_{pl}$ on the gate voltage. Moreover, the particular dependence of $\Delta U_{mis}$ on the positive gate voltages may be used to discriminate the contribution of each mechanism to the total THz detection. In fact, although a more quantitative analysis would require a precise estimation of the C factor in $\Delta U_{pl}$ and thermopower in $\Delta U_{TE}$, the small sign switch of $\Delta U_{mis}$ is significant and may be interpreted as an indication that the plasmonic mechanism was strongly weakened by the thermoelectric effect but still played an important role in our detector.

The detector figures of merit were calculated by following the procedure in Ref. 15. The responsivity R was estimated from the measured photovoltage assuming a full coupling of the radiation to the FET channel and detector active area equal to the diffraction limit. Therefore, only the lower limit of R was extracted. The maximum responsivity was achieved at 263 GHz and 325 GHz corresponding to ~0.25 V/W. Instead, R ~ 0.15 V/W was estimated at 295 GHz and 353 GHz. It is worth noting that the maximum responsivity was achieved for the frequencies where the highest temperature gradient was estimated, confirming a more efficient coupling of the THz radiation with the FET channel. Finally, due to the non-current-biased configuration, the thermal Johnson-Nyquist noise was assumed as the dominant contribution[14,16] to calculate the noise-equivalent-power (NEP). The minimum value of NEP was quantified as high as 80 nW/$\sqrt{Hz}$ and 160 nW/$\sqrt{Hz}$ at excitation of 263 GHz/325 GHz and 295 GHz/353 GHz, respectively.

In conclusions, we demonstrated room temperature THz detection with epitaxial graphene grown on silicon carbide substrate as the channel of an antenna-coupled field effect transistor. Photoresponse analysis revealed the presence of two independent detection mechanisms: plasmonic detection by means of the electron transport nonlinearity and thermoelectric effect ascribed to the presence of carrier density junction and induced temperature gradient across the FET channel. In fact, the behavior of the measured photovoltage as function of the gate voltage was qualitatively well explained in terms of plasma wave and thermoelectric photovoltages. Particularly, the photovoltages of both mechanisms were calculated by assuming the hydrodynamic transport model for the plasmonic regime and the Mott formulation for the TE effect. Their functional dependences as the gate voltage varies were found to be similar but with opposite sign, pointing out the competition of the two mechanisms. Specifically, the analysis of the experimental data suggested that the plasma wave detection may be the dominant mechanism, even though strongly counterbalanced by the TE response. The detrimental competition between the DS and TE effects resulted in NEP ~ 80 nW/$\sqrt{Hz}$ and photoresponsivity of about 0.25 V/W. This is one to two orders of magnitude lower than in plasmonic THz detectors fabricated by exfoliated graphene[15] or CVD graphene transferred on Si/SiO$_2$ substrate[16], respectively. The improvement of the plasmonic detector performance can be achieved with the maximization of the channel transconductance by minimizing the size of the ungated regions[16], and thus the associated series resistance, and using a quasi-free standing graphene obtained by hydrogen intercalation[27]. The latter would allow the graphene to be decoupled from the SiC substrate and the electrical properties to be improved[26], thus enhancing the plasmonic photoresponse of the detectors[44]. Although device improvements are required for a competitive photoresponse, these results demonstrated that epitaxial graphene grown on SiC is a promising candidate for compact scalable fast focal plane detectors for imaging applications.

**ACKNOWLEDGMENTS**

The work was supported by the European Union Seventh Framework Programme under grant agreement n° 604391 Graphene Flagship.

**FIGURES**

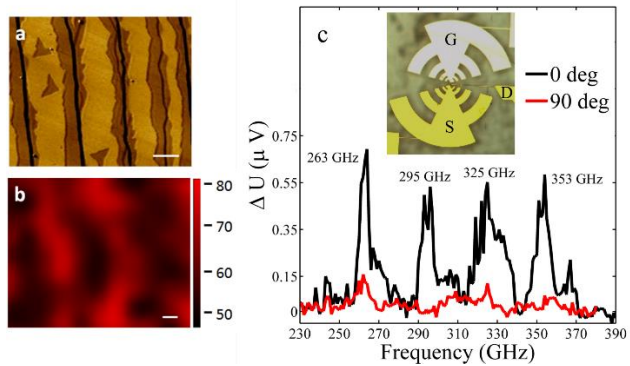

FIG. 1. Tapping-mode AFM phase signal (a) and Raman map of the 2D band (b) of graphene grown on SiC. The 2D peak width shows a bilayer character on the terraces and trilayer inclusions in the proximity of the SiC step. The scale bars are 1 µm. c) Photovoltage as a function of the incident radiation frequency measured at $V_g$ = 1.2 V, $V_{sd}$ = 0 mV and for two polarizations: parallel (0 degree) and orthogonal (90 degree) to the antenna axis. The data are not corrected by the spectral dependence of the source output power. The inset shows the optical image of the device.

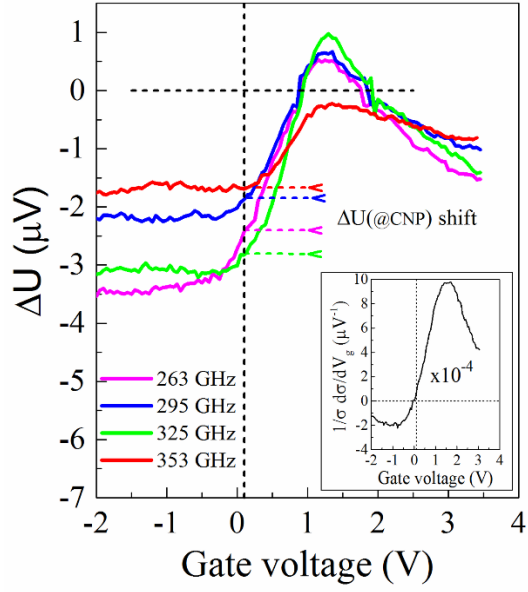

FIG. 2. Photovoltage measured as a function of the gate voltage at 263 GHz (magenta line), 295 GHz (blue line), 325 GHz (green line) and 353 GHz (red line) while keeping $V_{sd} = 0$ mV and polarization parallel to the antenna axis. The $\Delta U$ shift at CNP voltage from the 0 value is indicated for each photoresponse curve by arrow-dashed lines. The inset shows the gate voltage dependence of the photovoltage calculated as $1/\sigma\, d\sigma/dV_g$, where $\sigma$ is the source-drain conductivity of the GFET, for overdamped plasma waves-assisted detection. The dashed vertical and horizontal lines indicate the charge neutrality point and $\Delta U = 0$ V, respectively.

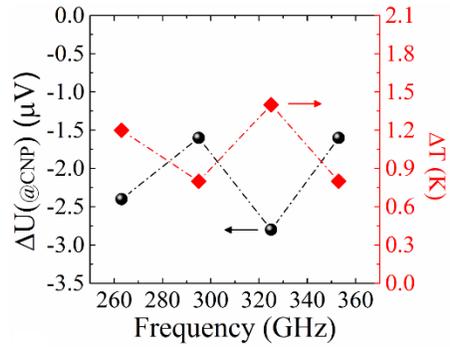

FIG. 3. ΔU at CNP (black dots) and the corresponding estimated temperature difference of the junction (red diamonds) as a function of the radiation frequency.

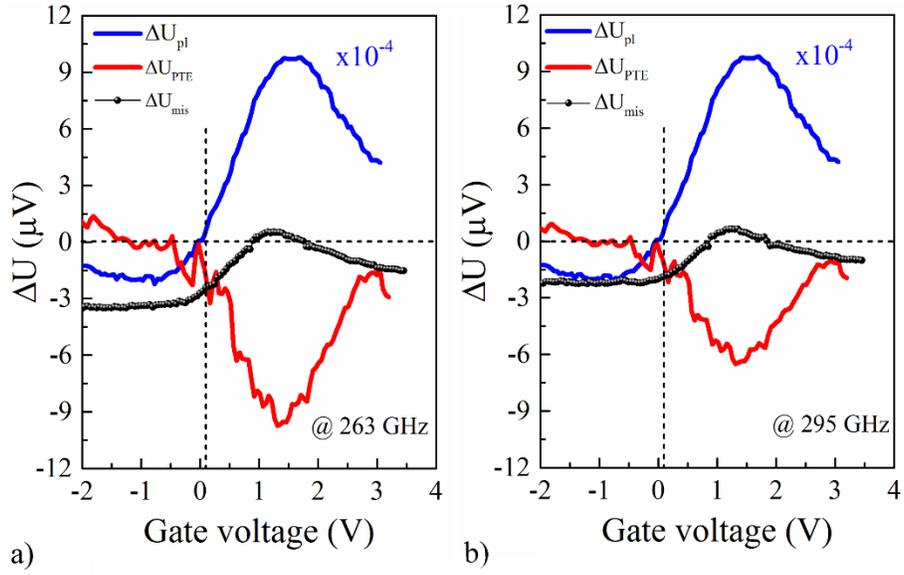

FIG. 4. Calculated plasmonic (blue lines), thermoelectric (red lines) and measured (black dots and lines) photo-induced voltage as a function of the gate voltage at 263 GHz (a) and 295 GHz (b) with $V_{sd} = 0$ mV and parallel polarization. The dashed vertical and horizontal lines indicate the charge neutrality point and $\Delta U = 0$ V, respectively.